\documentclass[aip,jcp,reprint,showkeys]{revtex4-1}
\pdfoutput=1
\usepackage{graphicx}
\usepackage{amsmath,amsthm}
\usepackage{verbatim}
\usepackage{enumitem}
\graphicspath{{./pics/}}

\begin{document}

\title{Self-Assembly of Patchy Colloidal Dumbbells}

\author{Guido Avvisati}
\email[]{g.avvisati@uu.nl}
\affiliation{Debye Institute for Nanomaterials Science, Utrecht University, Princetonplein 1, 3584CC Utrecht, The Netherlands}
\author{Teun Vissers}
\affiliation{SUPA, School of Physics and Astronomy, The University of Edinburgh, King’s Buildings, Peter Guthrie Tait Road, Edinburgh, EH9 3FD, United Kingdom}
\author{Marjolein Dijkstra}
\email[]{m.dijkstra1@uu.nl}
\affiliation{Debye Institute for Nanomaterials Science, Utrecht University, Princetonplein 1, 3584CC Utrecht, The Netherlands}

\date{\today}

\begin{abstract}
  We employ Monte Carlo simulations to investigate the self-assembly of patchy colloidal dumbbells interacting via a modified Kern-Frenkel potential by probing the system concentration and dumbbell shape. We consider dumbbells consisting of one attractive sphere with diameter $\sigma_1$ and one repulsive sphere with diameter $\sigma_2$ and center-to-center distance $d$ between the spheres. For three different size ratios, we study the self-assembled structures for different separations $l = 2d/(\sigma_1+\sigma_2)$ between the two spheres. In particular, we focus on structures that can be assembled from the homogeneous fluid, as these might be of interest in experiments. We use cluster order parameters to classify the shape of the formed structures. When the size of the spheres is almost equal, $q=\sigma_2/\sigma_1=1.035$, we find that, upon increasing $l$, spherical micelles are transformed to elongated micelles and finally to vesicles and bilayers. For size ratio $q=1.25$ we observe a continuously tunable transition from spherical to elongated micelles upon increasing the sphere separation. For size ratio $q=0.95$ we find bilayers and vesicles, plus faceted polyhedra and liquid droplets. Our results identify key parameters to create colloidal vesicles with attractive dumbbells in experiments.
\end{abstract}

\pacs{}
\keywords{Colloidal particles, self-assembly, computer simulations, Monte Carlo methods}

\maketitle

\section{\label{sec:intro}Introduction}
Colloidal self-assembly refers to the self-organisation process of nano- to micrometer-sized colloidal particles into larger structures \cite{bib:whitesides-self.assembly}. This process can be used for the fabrication of novel materials \cite{bib:kretzschmar-review.nanomaterials,bib:quintela-review.nanomaterials,bib:pine-colloidal.valence} and has potential applications in photonics \cite{bib:maldovan-diamond.crystals,bib:vanblaaderen-complex.colloids,bib:hynninen-photonic.crystal,bib:sawada-colloids.photonics} and medicine \cite{bib:lahann-biomaterials,bib:lahann-biphasic.janus,bib:delair-regenerative.medicine}. The ability to guide the self-assembly allows for a bottom-up approach to design and create specific materials. One way to achieve such guidance, is to engineer colloidal particles with discrete, attractive patches at well-defined locations on the surface of the particles \cite{bib:glotzer-anisotropic.assembly}. As several tuning parameters can in principle affect the self-assembly process, computer simulations provide an invaluable tool to explore the self-assembly of patchy particles models. 

One of the simplest patchy particle models is a sphere where one half is covered with an attractive patch that can interact with a similar patch on another sphere. The self-assembly of these so-called ``Janus'' particles has been investigated in computer simulations \cite{bib:solomon-review.colloids,bib:bianchi-review.patchy} and revealed the spontaneous formation of micelles and vesicles together with wrinkled sheets and different crystal structures \cite{bib:sciortino-janus, bib:hong-janus.clusters, bib:vissers-patchy.clusters, bib:vissers-janus.crystals}. In addition to Janus particles, past computer simulations studies have also investigated the effects of the number of patches and their surface distribution \cite{bib:sciortino-pacci.chains, bib:munao-pacci.tetra, bib:romano-pacci.tri.penta}, as well as the patch coverage fraction, patch shape and interaction range \cite{bib:romano-pacci.crystal, bib:romano-pacci.range, bib:vissers-tubes.lamellae} on the structure and the phase behaviour of patchy particles. On the experimental side, patchy particles can be synthesised in a large variety of shapes and with different patchy properties \cite{bib:imhof-colloids.buckling,bib:sacanna-review.colloids,bib:sacanna-review.patchy}. In some cases, the particles are already used to form complex ordered structures, e.g. clusters \cite{bib:hong-janus.clusters, bib:granick-charged.clusters, bib:solomon-spheroidal.patchy} and Kagome lattices \cite{bib:chen-pacci.triblock, bib:chen-kagome}. 

While many patchy particle models have been investigated so far, most systematic studies have focused on spherical colloids. Some studies have been performed on dumbbells with a selective attraction on one of the spheres. Experimentally, such dumbbells can be realised by introducing a variation in the surface roughness between the two spheres, such that the combination of electrostatic repulsion and depletion attraction leads to an effective attraction between the smooth spheres, which can result in the presence of a micellar fluid \cite{bib:kraft-dumbbells}. Previous theoretical works on attractive dumbbells have confirmed the existence of a micellar fluid and also reported bilayer formation \cite{bib:whitelam-peanuts}. Bilayer formation has also been observed in simulations for tangential hard dumbbells with tunable attraction strength \cite{bib:munao-patchy.dumbbells, bib:munao-janus.dumbbells}. However, spontaneous vesicles formation, which is found in molecular surfactants \cite{bib:israelachvili-colloid.science}, remains unobserved at the level of computer simulations of patchy colloidal dumbbells.

In this paper we use computer simulations to address the self-assembly of non-overlapping patchy dumbbells with an interaction range of half the diameter of the attractive sphere.  This is considerably longer than the interaction range in some experiments with depletion interactions \cite{bib:kraft-dumbbells}, but might still be realistic for e.g. nanoparticles or other types of interactions. We show that there are regions in our parameter space where we observe vesicle formation. Furthermore, we also observe structures which were previously reported, such as bilayers and micelles. The formation of the different structures is achieved by varying the size ratio and the sphere separation of the dumbbells, as well as the volume fraction of the system.

\section{\label{sec:model}Model and Methods}
In this section, we introduce the model definition and the method used to address the self-assembly of dumbbell-shaped particles. The dumbbells are understood to interact with one another and hereafter are referred to as ``attractive'' dumbbells. The details on how a single dumbbell particle is built and how the attractive potential is defined are found in subsection \ref{ssec:potential}, while the details on the simulation and analysis methods follow in subsections, respectively, \ref{ssec:mc.sims} and \ref{ssec:order.parameters}.

\subsection{\label{ssec:potential}Interaction Potential}
\begin{figure}[htb]
  \includegraphics[scale=2.5]{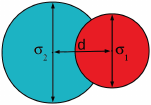}
  \caption{{\small Graphical representation of the geometry of one patchy dumbbell particle.}}
  \label{fig:db_geometry}
\end{figure}

Each dumbbell is assembled as follows: one attractive sphere of diameter $\sigma_1$ (red sphere in Fig. \ref{fig:db_geometry}) is placed at distance $d$ from another non-interacting sphere of diameter $\sigma_2$ (blue sphere in Fig. \ref{fig:db_geometry}) which acts as a steric constraint. Thus, the geometrical parameters are the size ratio, $q = \sigma_2/\sigma_1$, and the dimensionless sphere separation, $l = 2d/(\sigma_1+\sigma_2)$. When $l=0$ we simulate a Janus dumbbell, while for $l=1$ we obtain a dumbbell consisting of tangent spheres.
Furthermore, each dumbbell carries a normalised orientation vector, $\hat{\varepsilon}$, pointing from the center of the non-interacting sphere towards the attractive sphere. To describe the interactions between the patchy dumbbells, we employ an extension of the Kern-Frenkel potential \cite{bib:frenkel-kf.potential} for spherical patchy particles. The Kern-Frenkel potential consists of a spherically symmetric square-well potential modulated by an angular function which depends on the orientations of the particles. Given two dumbbells $i$ and $j$ with orientations $\hat{\varepsilon}_i$ and $\hat{\varepsilon}_j$ respectively, their pair potential reads:
\begin{equation}
  u_{ij}=
  u^{\mathrm{SW}}(r^{11}_{ij})f(\hat{r}^{11}_{ij},\hat{\varepsilon}_i,\hat{\varepsilon}_j) + u^{\mathrm{HS}}_{ij}(r^{11}_{ij},r^{12}_{ij},r^{21}_{ij}, r^{22}_{ij})
  \label{eq:interaction.energy}
\end{equation}
where $\hat{r}^{11}_{ij}$ is the normalised vector connecting the attractive spheres of dumbbells $i$ and $j$, $r^{11}_{ij}=\left|\boldsymbol{r}^1_j-\boldsymbol{r}^1_i\right|$ denotes the absolute center-of-mass distance between the attractive hard spheres of dumbbells $i$ and $j$, $r^{12}_{ij}$ the distance between the attractive hard sphere on dumbbell $i$ and the non-interacting hard sphere on dumbbell $j$, $r^{21}_{ij}$ the distance between the non-interacting hard sphere on dumbbell $i$ and the attractive hard sphere on dumbbell $j$, and $r^{22}_{ij}$ the distance between the non-interacting hard spheres on dumbbells $i$ and $j$. The square-well interactions between the attractive spheres on the two dumbbells are given by:
\begin{equation}
  \beta u^{\mathrm{SW}}(r^{11}_{ij})= 
  \begin{cases}
    \beta\epsilon & \mathrm{for}\; \sigma_1 \leq 
    r^{11}_{ij} < \sigma_1 + \Delta \\
    0 & \mathrm{for}\; r^{11}_{ij} \geq \sigma_1 + \Delta
  \end{cases} 
  \label{eq:radial.part}
\end{equation}
where $\beta=1/k_BT$ is the inverse temperature, $k_B$ the Boltzmann's constant, $\epsilon < 0$ and $\Delta$ are the square-well (SW) parameters representing, respectively, the interaction strength and range. The orientational Kern-Frenkel part gives directionality to the interaction potential and is given by:
\begin{equation}
  f(\hat{r}^{11}_{ij},\hat{\varepsilon}_i,\hat{\varepsilon}_j)=
  \begin{cases}
    1 & \mathrm{if}\; 
    \begin{cases} 
      & \hat{\varepsilon}_i \cdot \hat{r}^{11}_{ij} 
      \geq \cos\,\delta \\ 
      \mathbf{and} & \hat{\varepsilon}_j 
      \cdot \hat{r}^{11}_{ji}
      \geq \cos\,\delta \\ 
    \end{cases} \\
    0 & \mathrm{otherwise}
  \end{cases} 
  \label{eq:angular.part}
\end{equation}    
The opening angle $\delta$ depends on the geometry of the particle via the relation:
\begin{equation}
  \cos\delta = \frac{1}{4d\sigma_1}\cdot\left(\sigma_2^2 - \sigma_1^2 - 4d^2\right)
  \label{eq:trigon}
\end{equation}
and follows from trigonometry by connecting the centre of mass of the small attractive sphere to the intersection point between the small attractive sphere and the larger non-interacting sphere. As a consequence, the attractive spheres on two different dumbbells can not interact with each other through the volume of the non-interacting spheres. Finally, the hard-sphere part of the potential assures that two dumbbells do not overlap with each other and reads:
\begin{equation}
  u^{\mathrm{HS}}_{ij}(r^{11}_{ij},r^{12}_{ij}, r^{21}_{ij}, r^{22}_{ij}) =
  \begin{cases}
    \infty & \mathrm{if}\; 
    \begin{cases} 
      & r^{11}_{ij} < \sigma_{1} \\ 
      \mathbf{or} & r^{12}_{ij} < (\sigma_{1}+\sigma_{2})/2 \\
      \mathbf{or} & r^{21}_{ij} < (\sigma_{1}+\sigma_{2})/2 \\ 
      \mathbf{or} & r^{22}_{ij} < \sigma_{2} \\ 
    \end{cases} \\
    0 & \mathrm{otherwise}
  \end{cases} 
  \label{eq:hs.part}
\end{equation}

\begin{figure}[htb]
  \includegraphics[scale=0.15]{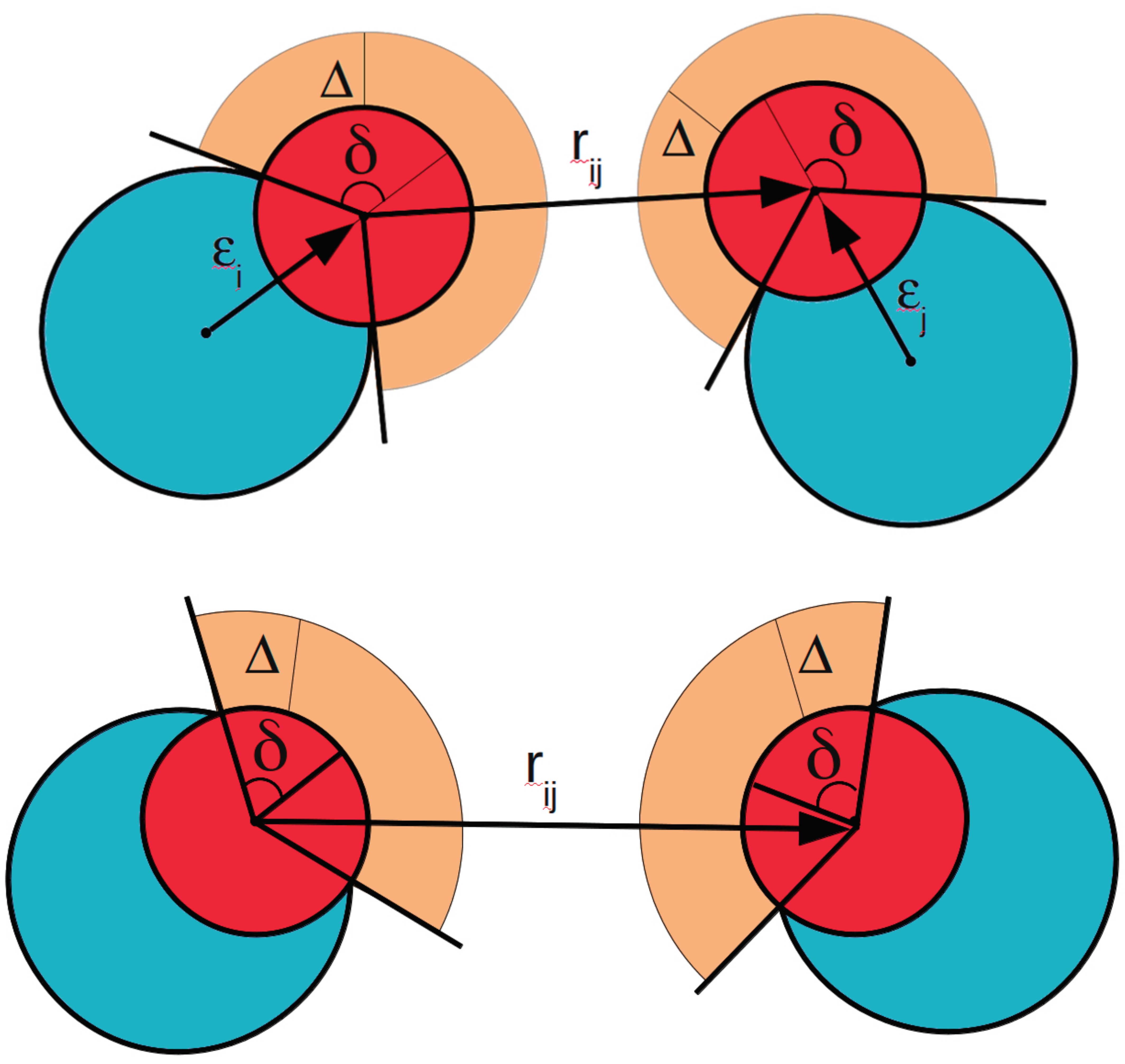}
  \caption{{\small Graphical representations of the interaction between a pair of patchy dumbbells in two different cases. Top panel: dumbbells with size ratio $q=\sigma_2/\sigma_1=1.25$ and dimensionless sphere separation $l=2d/(\sigma_1+\sigma_2) > 1/3$, corresponding to $\delta > 90^{\circ}$. Bottom panel: dumbbells with size ratio $q=\sigma_2/\sigma_1=1.25$ and dimensionless sphere separation $l=2d/(\sigma_1+\sigma_2) < 1/3$, corresponding to $\delta < 90^{\circ}$ (bottom panel). The attractive spheres are denoted with red, the non-attractive spheres are denoted with blue. The orange area represents the interaction range.}}
  \label{fig:db_interaction}
\end{figure}

\noindent
A representation of the interaction model is given in Fig. \ref{fig:db_interaction}, for two different values of $\delta$. Note that the attractive region (in orange) does not intersect with the non-interacting spheres.

\subsection{\label{ssec:mc.sims}Monte Carlo Simulations}
We perform Monte Carlo simulations in the canonical ensemble (MC-$NVT$) with $N=1024$ dumbbells in a volume $V$ and temperature $T$ with cubic periodic boundary conditions. We define the volume fraction $\phi=\rho V_{DB}$ inside the simulation box, where $\rho=N/V$ denotes the total number density, and $V_{DB}$ is the volume of a single dumbbell. We employ single particle translation and rotation moves \cite{bib:frenkel-ums} to explore the configurational phase space. The simulations are $2-6 \times10^7$ MC steps long, where a single MC step is defined as $N$ attempted moves (either translations or rotations). 

To be more precise, our model possesses a five-fold parameter space $\{\phi, \beta\varepsilon, \Delta/\sigma_1, l, q\}$, denoting the volume fraction $\phi$, the well depth $\beta\varepsilon$, the interaction range $\Delta/\sigma_1$, the dimensionless distance between the two spheres in the dumbbell $l$ and the size ratio between the non-interacting and the attractive sphere $q$. Throughout this work, we fix the interaction strength to $\beta\varepsilon=-3.58$, which is sufficient to observe self-assembly \cite{bib:munao-janus.dumbbells}. For all our simulations, we set the interaction range $\Delta=0.5\sigma_1$, to half the diameter of the attractive sphere in the dumbbell, similar to the interaction range used previously to study Janus particles \cite{bib:sciortino-janus} and patchy dumbbells \cite{bib:munao-patchy.dumbbells, bib:munao-janus.dumbbells}.

\subsection{\label{ssec:order.parameters}Order Parameters}
\begin{figure}[htb]
  \includegraphics[scale=0.35]{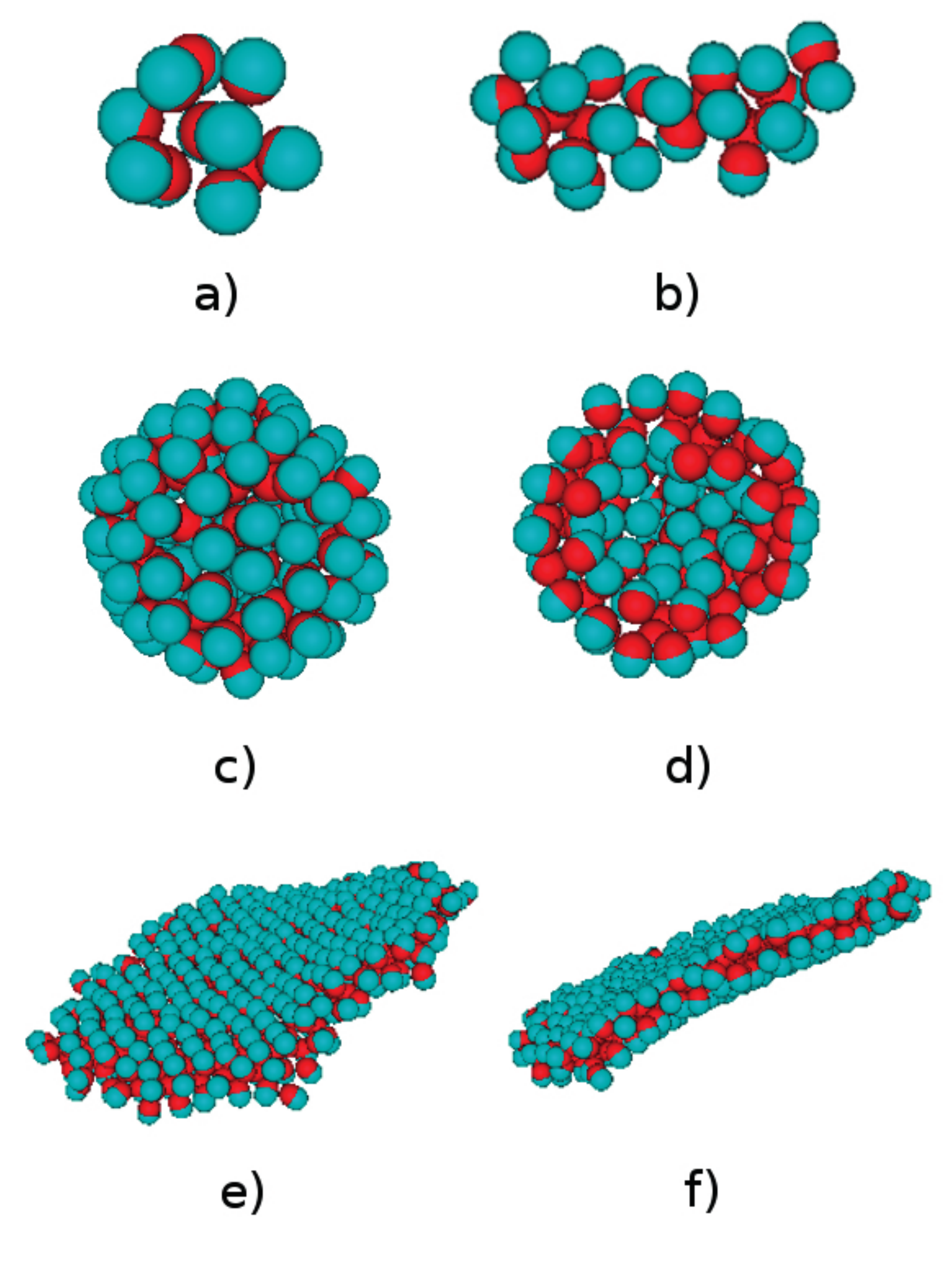}
  \caption{{\small Most common self-assembled structures found for patchy dumbbells with size ratio $q=1.035$ at the end of the simulation runs: a) spherical micelle $l=0.08$, b) elongated micelle $l=0.13$, c) vesicle $l=0.20$, d) cut-through of a vesicle, e) and f) bilayers $l=0.28$.}}
  \label{fig:simstructures}
\end{figure}

To compare the outcome of computer simulations for different parameters, we choose a systematic approach to analyse the final configuration of a simulation. To this end, we employ a cluster analysis method and use three order parameters to classify the clusters. This approach is similar to the one used in Ref. \cite{bib:whitelam-peanuts}, except here we introduce an additional order parameter in order to deal with the additional encountered structures. In Fig. \ref{fig:simstructures} we show the typically encountered aggregate shapes, found for patchy dumbbells with size ratio $q=1.035$: a spherical micelle (a), an elongated micelle (b), a vesicle (c,d), and a bilayer (e,f).

\noindent
The procedure to identify and classify clusters is the following. First, we identify  particles as interacting neighbours if they have a mutual bond, i.e. they attract each other according to Eq.~\ref{eq:interaction.energy}. Then, a cluster is defined as a contiguous set of neighbouring particles. For each cluster found in the final configuration of a simulation run, we compute three cluster order parameters $\mathcal{M}$, $\mathcal{B}$ and $\mathcal{V}$ defined as
\begin{align}
  \mathcal{M} & = \frac{1}{N_c}\sum_{i=1}^{N_c}\cos\theta_i \label{eq:micelles} \\
  \mathcal{B} & = \frac{2}{N_c(N_c-1)}\sum_{(ij)}\left(\hat{\varepsilon}_i\cdot\hat{\varepsilon}_j\right)^2 \label{eq:bilayers} \\
  \mathcal{V} & = \frac{1}{N_c}\sum_{i=1}^{N_c}\left(1-\sin\theta_i\right) \label{eq:vesicles}
\end{align}
where $N_c$ is the number of dumbbells in the cluster and $\sum_{(ij)}$ denotes the sum over all particle pairs in a cluster. The quantity $\cos\theta_i$ is defined as
\begin{equation}
  \cos\theta_i = \hat{\varepsilon}_i\cdot\frac{\boldsymbol{r}_{cm}-\boldsymbol{r}_i}{\left|\boldsymbol{r}_{cm}-\boldsymbol{r}_i\right|} 
  \label{eq:theta} 
\end{equation}
with $\boldsymbol{r}_{cm}$ denoting the center of mass of the cluster, $\boldsymbol{r}_{i}$ indicating the center of mass of dumbbell $i$ and $\hat{\varepsilon}_i$ labelling the orientation of dumbbell $i$.
For a perfectly spherical micelle we have $\mathcal{M} = 1$, whereas $\mathcal{B} = 1$ for a bilayer, i.e. a collection of dumbbells that are aligned either perfectly parallel or anti-parallel with respect to each other. To further discriminate the structures, $\mathcal{V}$ detects whether particles are not oriented perpendicular with respect to the vector connecting them to the center of mass of the clusters. For infinitely long and flat bilayers, $\mathcal{V}$ would be low as the particles are oriented orthogonally to the vector connecting them to the center of mass of the cluster. 

We found the following criteria appropriate to classify the shape of the clusters (more details can be found in the Appendix \ref{apx:class.details}).
\begin{enumerate}
\item[{\includegraphics[scale=1.3]{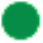}}] Spherical Micelle: $\mathcal{M} \geq 0.9$. 
\item[{\includegraphics[scale=1.2]{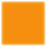}}] Elongated Micelle: $\mathcal{M} < 0.9$ and $\mathcal{B} < 0.4$. 
\item[{\includegraphics[scale=1.3]{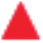}}] Vesicle: $\mathcal{M} < 0.5$ and $\mathcal{V} \geq 0.3$. 
\item[{\includegraphics[scale=1.3]{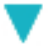}}] Bilayer: $\mathcal{M} < 0.5$ and $\mathcal{B} \geq 0.4$.
\item[{\includegraphics[scale=1.1]{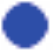}}] Liquid Droplet: $\mathcal{M} < 0.5$ and $0.3 \leq \mathcal{V} \leq 0.5$.
\item[{\includegraphics[scale=1.3]{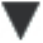}}] Other: aggregates not belonging to any previous category. 
\end{enumerate}
\noindent
where the symbols will be used later. According to this classification, clusters cannot fall under multiple categories. Furthermore, when two or more symbols are found together, it means that at least 20\% of the clusters belonged to the corresponding category. 

Although we rely on the automated classification method, we also use direct visual inspection as a consistency check. If we find disagreement between the two methods, visual inspection will be preferred. This case happens rarely and it will be explicitely mentioned in the captions of the following figures.

Using the depicted classification scheme, we can now map out, for varying size ratio $q$, the self-assembly state diagrams for patchy dumbbells in the sphere separation $l$--volume fraction $\phi$ representation. 

\section{\label{sec:results}Results}
We have investigated the self-assembly of patchy dumbbells for three different size ratios, $q=1.035$, $q=1.25$ and $q=0.95$. For each case, we have classified the structures according to the order parameters introduced in section \ref{ssec:order.parameters}.

\begin{figure}[htb]
  \includegraphics[scale=0.12]{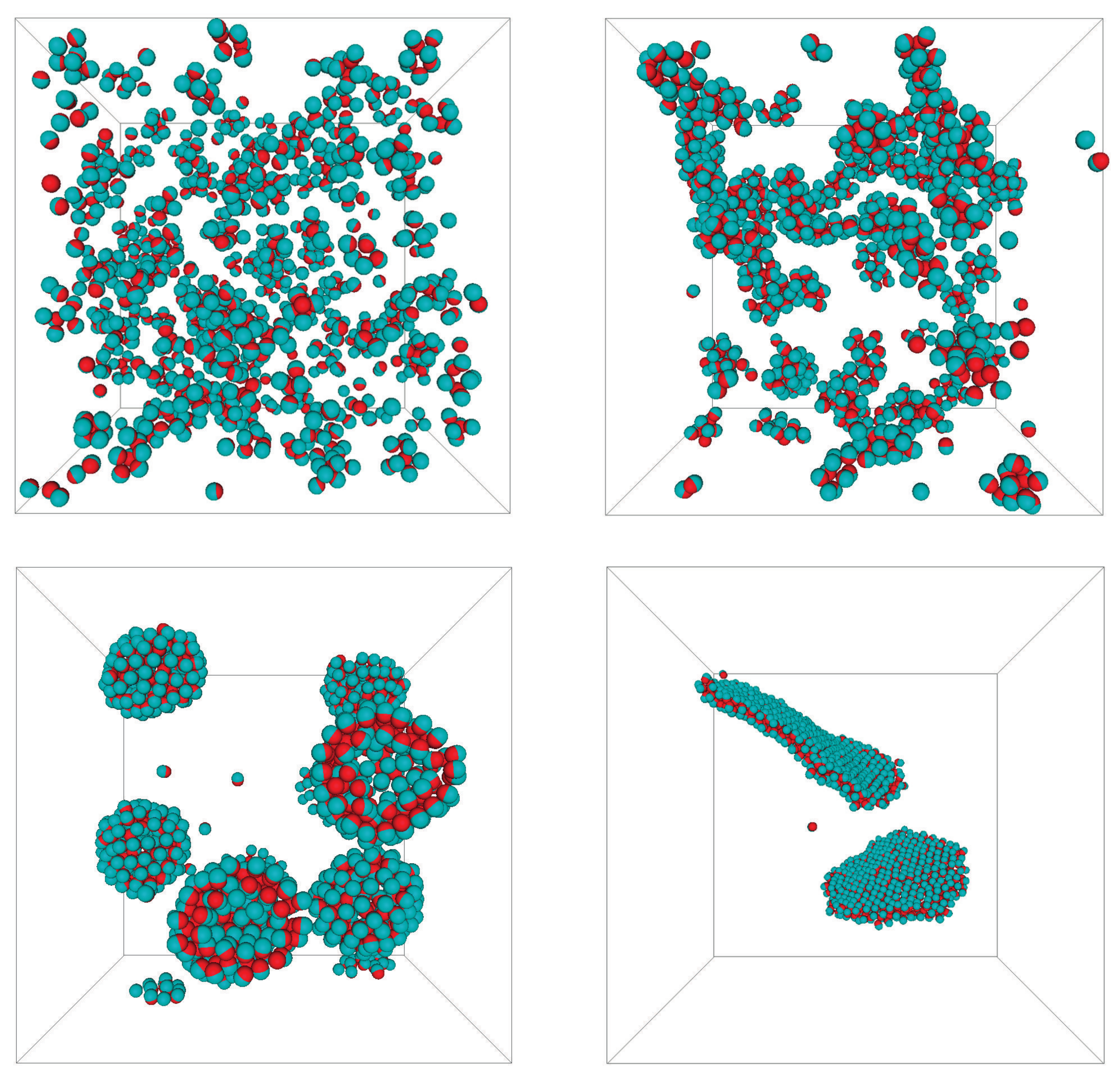}
  \caption{{\small Typical simulation snapshots for patchy dumbbells with size ratio $q=1.035$ for different values of volume fraction $\phi$ and sphere separation $l$. Top left: spherical micelles at $(\phi=0.03,l=0.05)$. Top right: elongated micelles at $(\phi=0.03,l=0.14)$. Bottom left: vesicles at $(\phi=0.035,l=0.175)$. Bottom right: bilayers at $(\phi=0.008,l=0.34)$.}}
  \label{fig:db.snap}
\end{figure}

For $q=1.035$, we observe a remarkably rich self-assembly behaviour, including the formation of micellar structures ranging from spherical to non-spherical shape. In addition, we also find vesicles and bilayers (see Fig.~\ref{fig:db.snap}). Note that for the micelles (Fig.~\ref{fig:db.snap} top panels) the attractive spheres point inwards, whereas for the vesicles and the bilayers the dumbbells form a double-layer where a part of the attractive spheres points inwards and the other part of them outwards.
\begin{figure*}[htb]
  \includegraphics[scale=0.55]{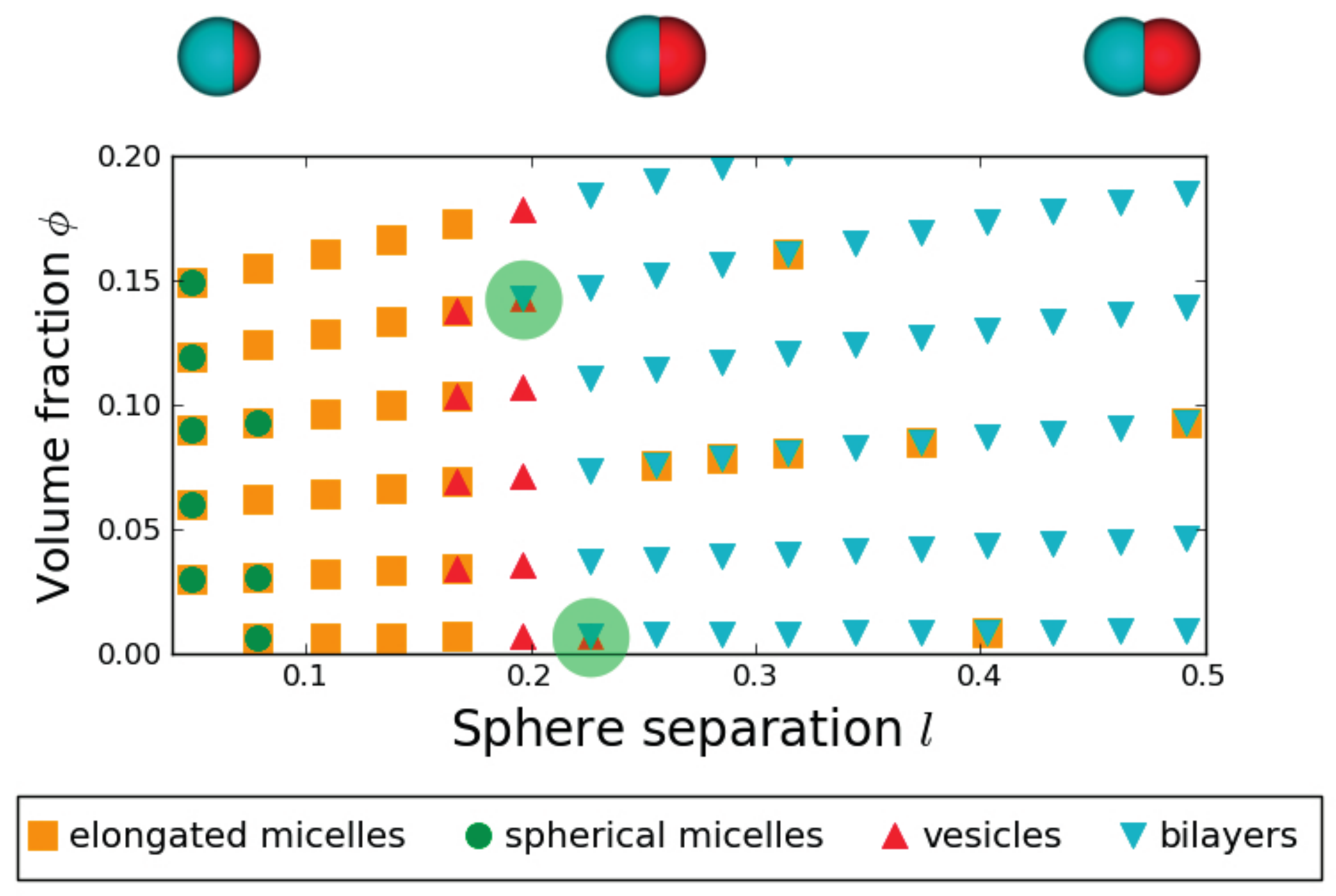}
  \caption{{\small State diagram of patchy colloidal dumbbells for size ratio $q=1.035$, interaction strength $\beta\varepsilon=-3.58$ and interaction range $\Delta=0.5\sigma_1$ in the sphere separation $l=2d/(\sigma_1+\sigma_2)$-volume fraction $\phi$ representation. The green-shaded circles indicate the state points chosen to investigate the stability of the vesicles with respect to the bilayers.}}
  \label{fig:diagram.1035}
\end{figure*}

The self-assembly state diagram for patchy dumbbells with size ratio $q=1.035$ as a function of volume fraction $\phi$ and dimensionless sphere separation $l$ is shown in Fig. \ref{fig:diagram.1035}. We find that the transition from one regime to the other is fully determined by the sphere separation $l$: micelles are observed for small separations $l$ when the attraction is more directional, whereas bilayers are found for large separations $l$, i.e., when the patchy interaction is less directional and the steric constraint by the non-interacting sphere of the dumbbell becomes more apparent. For intermediate separations, we find vesicles which are favoured due to a delicate balance between the directionality of the attraction and the geometric anisotropy of the particle.

For two state points, $[(l=0.23,\phi=0.007),(l=0.19,\phi=0.14)]$ we find that both bilayers and vesicles form in the simulation box. On increasing simulation time, we observe that the vesicles are not stable when the size of the cluster is well below $N_{c}=90$ and that in this case they open up to become small bilayer sheets. On the contrary, if the size of the vesicles is above $N_{c}=90$ they do not break up. However, as this behaviour is inferred by analysing configurations in the MC-$NVT$ simulations, it might be interesting to check the observation with explicit free-energy calculations using grand-canonical Monte Carlo simulations in single clusters as used in Refs.~\cite{bib:kraft-dumbbells, bib:vissers-janus.crystals}. To understand the relevance of the observed structures, it is worth mentioning that vesicles can be employed as drug containers in drug delivery processes \cite{bib:goymann-drug.vesicles,bib:cevc-drug.vesicles,bib:vohra-drug.nano}, while the bilayers offer the possibility of building large two-dimensional colloidal structures from very simple building blocks, which can be useful for application in photonics \cite{bib:zheng-bilayer.photonics}. 

\begin{figure*}[htb]
  \includegraphics[scale=0.78]{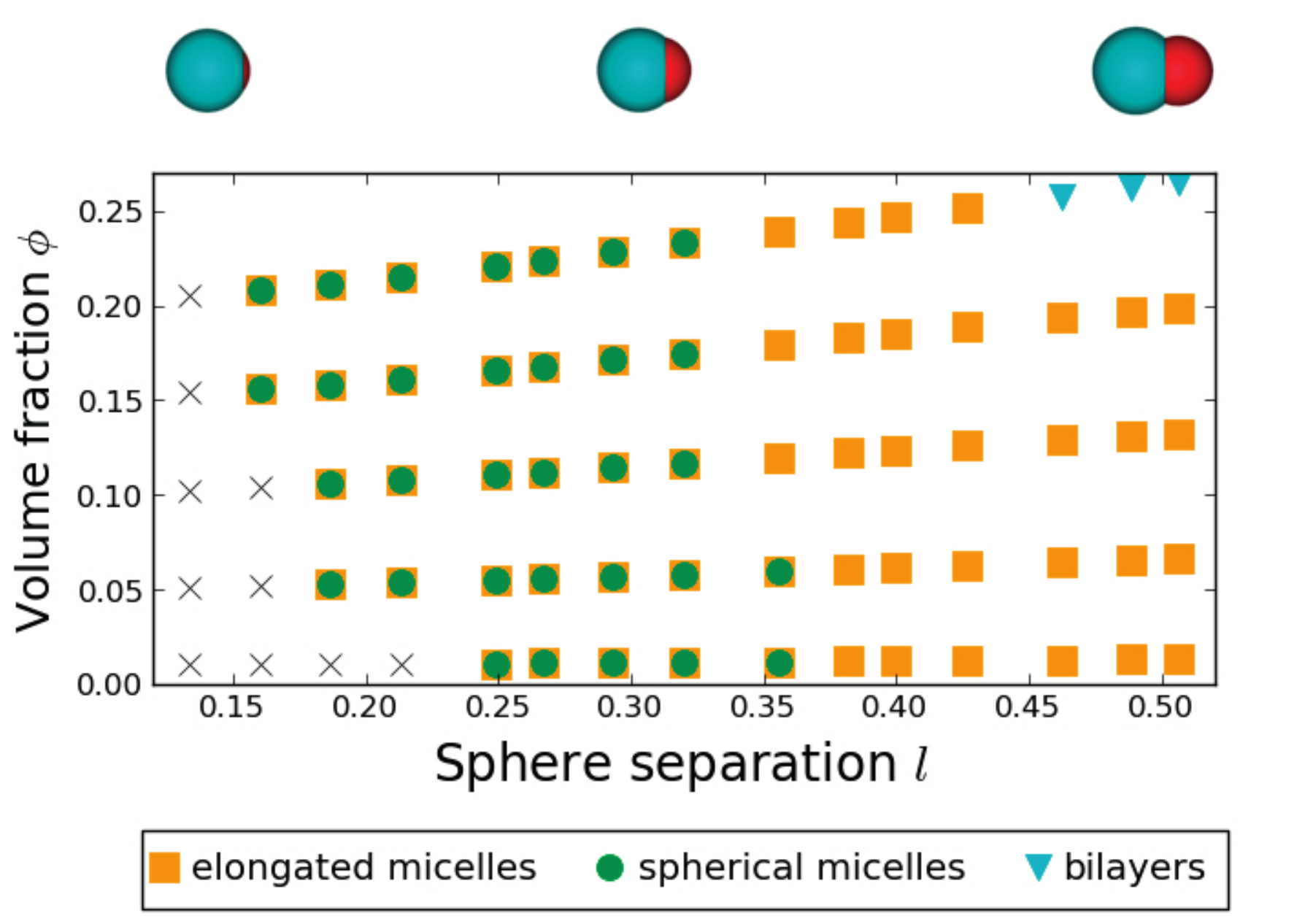}
  \caption{{\small State diagram of patchy colloidal dumbbells for size ratio $q=1.25$, interaction strength $\beta\varepsilon=-3.58$ and interaction range $\Delta=0.5\sigma_1$ in the sphere separation $l=2d/(\sigma_1+\sigma_2)$-volume fraction $\phi$ representation. The crosses indicate the regime where aggregation is not relevant.}}
  \label{fig:diagram.1250}
\end{figure*}
As we turn to size ratio $q=1.25$ for which the state diagram is shown in Fig. \ref{fig:diagram.1250}, we observe that, with respect to the size ratio $q=1.035$, the micellar region has grown at the expense of the vesicle and bilayer regime. Additionally, for small sphere separations we observe a regime where the total number of aggregated particles is smaller than the number of free monomers (denoted with crosses in the figure). The size of this regime decreases upon increasing the volume fraction. The bilayers can only be found at high volume fractions and high sphere separations. However, we do observe elongated micelles which have bilayer-like characteristics, close to the points where bilayers are observed. 

\begin{figure*}[htb]
  \includegraphics[scale=0.6]{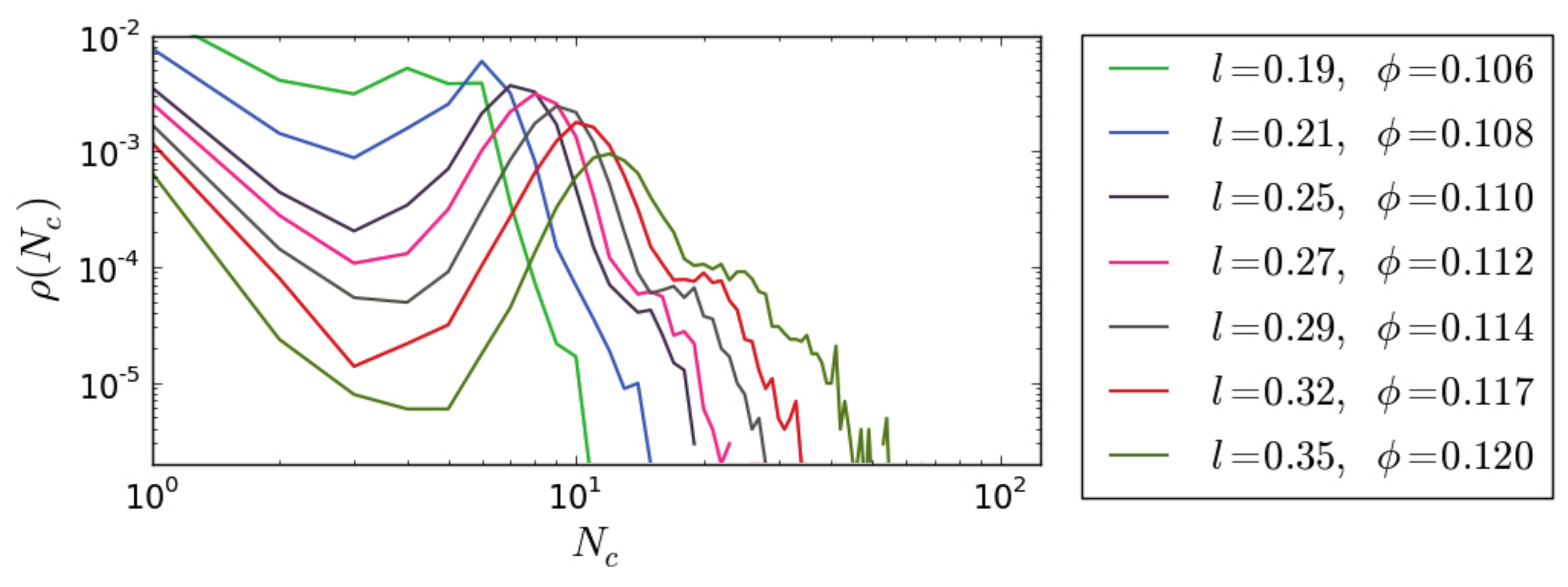}
  \caption{{\small Cluster size distribution $\rho(N_c)\equiv n_{N_c}/V$ as a function of the number of particles $N_c$ in a cluster for patchy colloidal dumbbells with size ratio $q=1.25$, interaction strength $\beta\varepsilon=-3.58$, interaction range $\Delta=0.5\sigma_1$, and varying sphere separation $l$ and volume fraction $\phi$ as labelled.}}
  \label{fig:csd.1250}
\end{figure*}

We additionally compute the cluster size distribution for $q=1.25$ for different values of sphere separation $l$, but all at the same number density $\rho\sigma_{1}^3=0.1$ (note however that the volume slightly changes as the two spheres of the dumbbell become more separate upon increasing $l$). This is defined as number of clusters with size $N_c$, $n_{N_c}$, divided by the box volume $V$. The cluster size distributions as shown in Fig. \ref{fig:csd.1250} for patchy dumbbells with a size ratio $q=1.25$ are strongly peaked, with the peak shifting to higher cluster sizes upon increasing the sphere separation $l$. Indeed, in this regime where micellar clusters form, the radius of curvature of a cluster becomes larger as the particles become more elongated, in turn allowing the clusters to grow larger. It is interesting to compare our results to Ref.~\cite{bib:whitelam-peanuts}. The size ratio here considered, $q=1.25$, is the closest value we have to the lowest value considered in that work ($q=1.4$), while their sphere separation in terms of $l=2d/(\sigma_1+\sigma_2)$ reads $l=0.53$. For these conditions, our estimate of the average cluster size $\left\langle N_c\right\rangle$ is $30$ times higher than the one in Ref.~\cite{bib:whitelam-peanuts}. The reason for this large difference is two-fold: firstly, the size ratio considered in Ref.~\cite{bib:whitelam-peanuts} is already $12\%$ larger than ours, secondly -- and more importantly -- the interaction range considered in this work ($\Delta=0.5\sigma_1$) is three times as large as the one considered in Ref.~\cite{bib:whitelam-peanuts}. Indeed, when we performed simulations with $q=1.4$ and shorter interaction range ($\Delta=0.15\sigma_1$) similar values were obtained for the average cluster size, being $\left\langle N_c\right\rangle\sim 9$ in this work and $\left\langle N_c\right\rangle\sim 7$ in Ref. \cite{bib:whitelam-peanuts}.

For size ratio $q=1.035$ and $q=1.25$, and for small values of the sphere separation $l$, a comparison can be made with the case of spherical patchy particles with low surface coverage \cite{bib:munao-low.coverage}. In view of this context, the size ratio $q=1.035$ and the sphere separation $l<0.1$ would correspond to a patch surface coverage of $\chi=0.4$. 
In this case, our model and the one in Ref. \cite{bib:munao-low.coverage} yield the same self-assembled structures, i.e. micelles, on all the volume fractions investigated in this work. 

\begin{figure*}[htb]
  \includegraphics[scale=0.55]{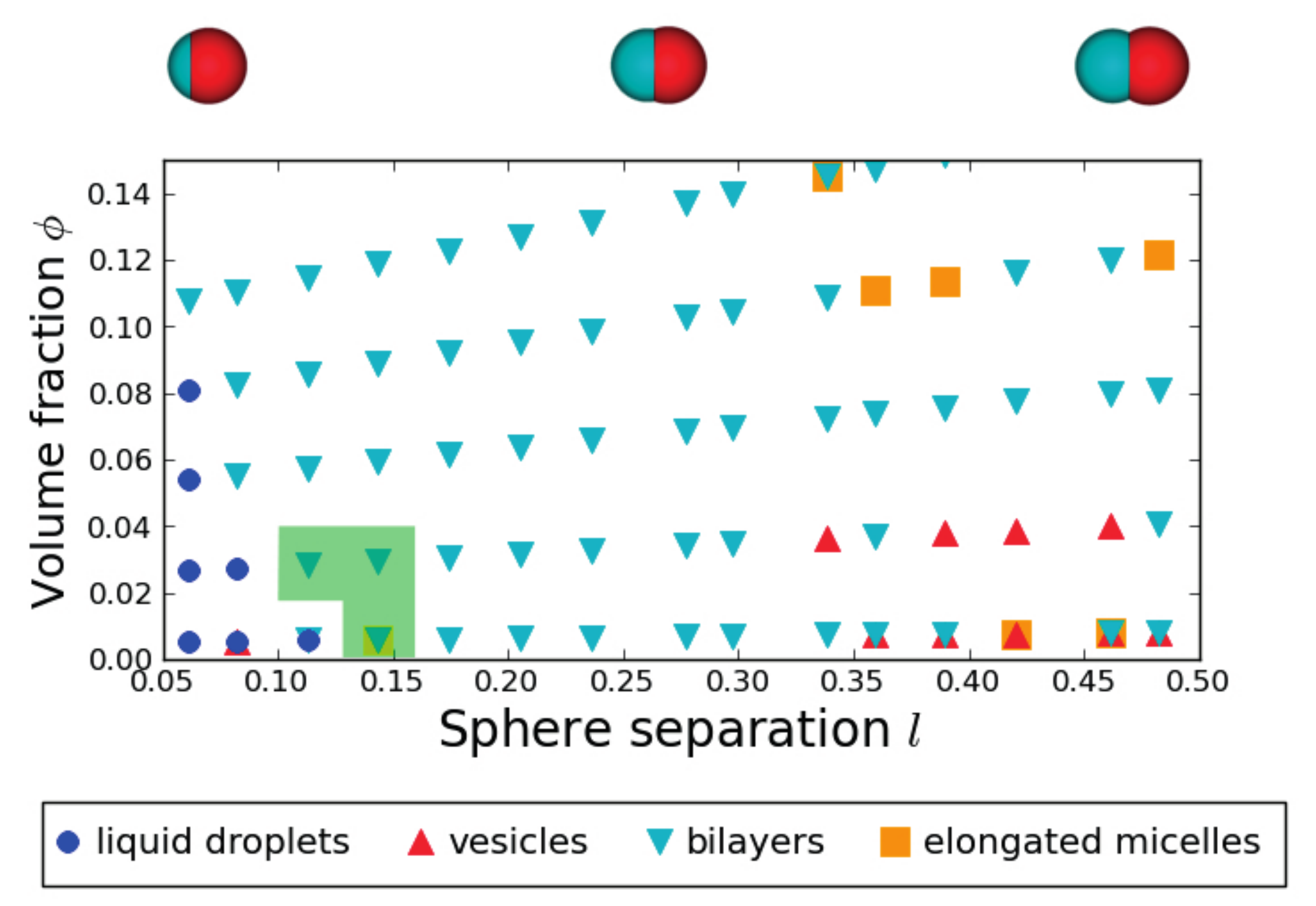}
  \caption{{\small State diagram of patchy colloidal dumbbells for size ratio $q=0.95$, interaction strength $\beta\varepsilon=-3.58$ and interaction range $\Delta=0.5\sigma_1$ in the sphere separation $l=2d/(\sigma_1+\sigma_2)$-volume fraction $\phi$ representation. The green-shaded region signals, for the state points inside it, the presence of faceted polyhedra in addition to the reported structures. The classification of five state points in the diagram has changed due to visual inspection.}}
  \label{fig:diagram.0950}
\end{figure*}

Finally, we have also investigated patchy dumbbells with size ratio $q=\sigma_2/\sigma_1=0.95$, where the non-interacting sphere is smaller than the attractive sphere. The corresponding state diagram for different values of $\phi$ and $l$ is given in Fig.~\ref{fig:diagram.0950}. The formation of bilayers is observed for most sphere separations. For larger separations, the formation of vesicles is observed. Note that this region is located at a different range of separations with respect to the size ratio $q=1.035$. While the size of the vesicles for size ratio $q=0.95$ is up to 10 times larger than the one found for $q=1.035$, both fall into the correct classification category, given by the $\mathcal{V}$ order parameter. 
\begin{figure*}[htb]
  \includegraphics[scale=0.7]{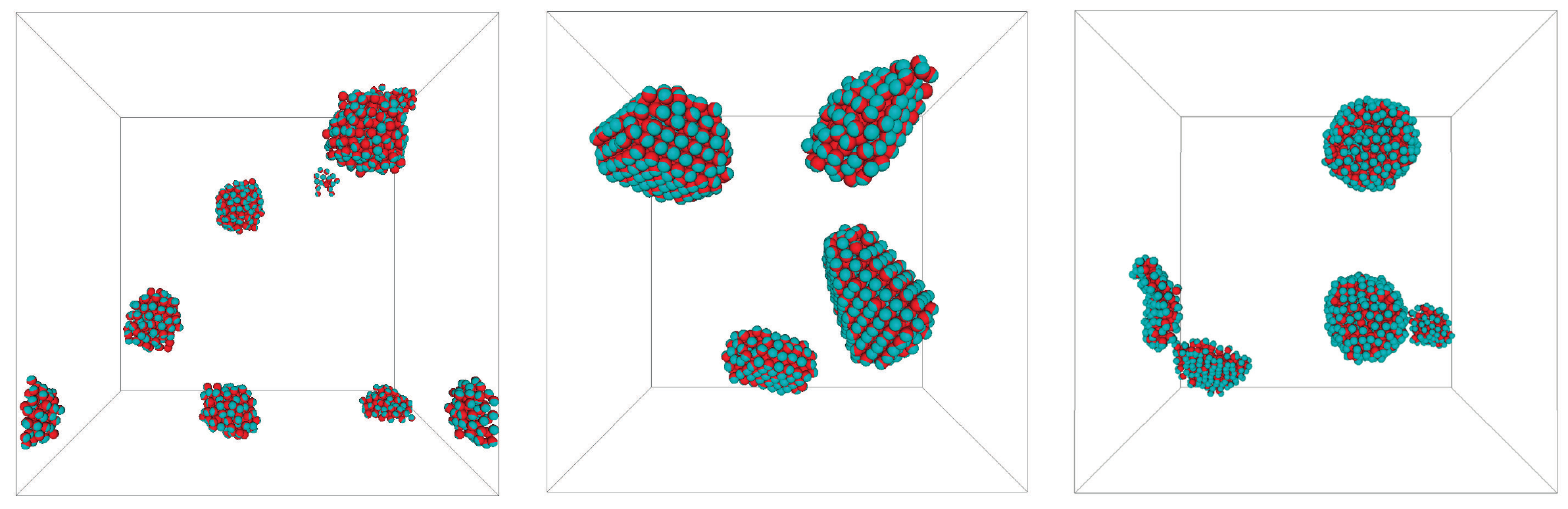}
  \caption{{\small Typical final configurations for patchy colloidal dumbbells with a size ratio $q=0.95$, interaction strength $\beta\varepsilon=-3.58$, interaction range $\Delta=0.5\sigma_1$. Left panel: Liquid droplets at packing fraction $\phi=0.005$ and sphere separation $l=0.06$. Middle panel: faceted polyhedra at packing fraction $\phi=0.028$ and sphere separation $l=0.08$. Right panel: vesicles at packing fraction $\phi=0.008$ and sphere separation $l=0.46$.}}
  \label{fig:VEandFP}
\end{figure*}

Interestingly, we also observe the formation of faceted polyhedra (see Fig.~\ref{fig:VEandFP}), which were also found in Ref.~\cite{bib:whitelam-peanuts,bib:cacciuto-janus}. In addition, we observe the formation of liquid droplets \cite{bib:cacciuto-janus} -- disordered, liquid-like, aggregates of particles which is to be expected as in the limit $l\rightarrow 0$ and $q\rightarrow 1$, the system reduces to a square-well fluid at a state point that lies well-inside the two-phase gas-liquid coexistence region \cite{bib:lago-sw.vl.equilibrium}.

\begin{figure*}[htb]
  \includegraphics[scale=0.8]{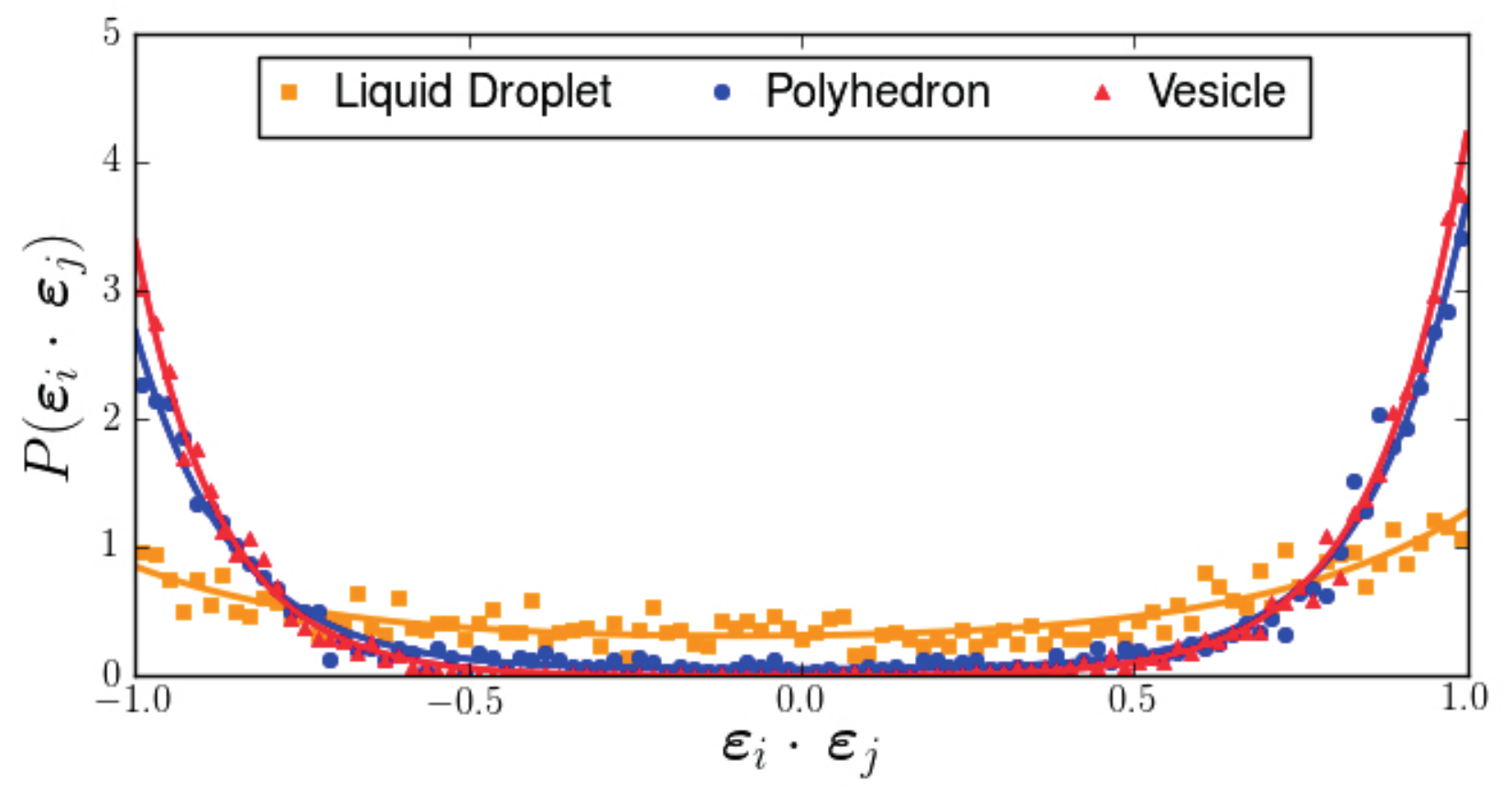}
  \caption{{\small Orientational probability distribution function for the representative aggregate type: faceted polyhedron, vesicle and liquid droplet. While faceted polyhedra and vesicles show similar degree of orientational ordering of the neighbouring particles, the disordered liquid droplets exhibit a more isotropic distribution. The solid lines are least-square fits to a double exponential function.}}
  \label{fig:ipd}
\end{figure*}

From the order parameter analysis, liquid droplets and faceted polyhedra look very similar. To distinguish between them, we calculate the orientational probability distribution function $P(\hat{\varepsilon}_i\cdot\,\hat{\varepsilon}_j)$, for a set of neighbouring particles $i$ and $j$, all belonging to the same cluster (Fig \ref{fig:ipd}). While it is apparent that pairs of dumbbells inside faceted polyhedra and vesicles have a strong tendency to be aligned or counter-aligned with respect to each other, the liquid droplets display a more isotropic distribution as the dumbbells are oriented more randomly with respect to each other.

Upon increasing the sphere separation, at low volume fractions, we observe a transition from liquid droplets to faceted polyhedra to bilayers that is similar to what has been reported in a previous work on amphiphilic spherical Janus colloids \cite{bib:cacciuto-janus}. For small sphere separations, this comparison is justified, since here in both cases the particles are approximately spherical. However, as the sphere separation increases the shape of our dumbbells becomes very different from spheres, explaining why the vesicles are not found in Ref.~\cite{bib:cacciuto-janus}. 

Finally, comparison with the state diagrams for $q=1.035$ and $q=1.25$ suggests that bilayers are the most frequently encountered structures for size ratio $q=0.95$. For size ratio $q=1.035$ the bilayers already become less dominant at higher sphere separations, at the cost of other interesting structures such as vesicles and spherical micelles. For size ratio $q=1.25$ the bilayers are only present in a small range of volume fractions, leaving more room for the formation of elongated and spherical micelles.

\section{\label{sec:summary}Conclusions}
    We have performed Monte Carlo simulations on patchy colloidal dumbbells consisting of one hard attractive sphere and one hard non-interacting sphere. To model the patchy interactions between the attractive spheres, we have extended the Kern-Frenkel potential to the case of non-spherical particles. In particular, we have investigated the effect of varying the size ratio $q=\sigma_2/\sigma_1$ between the two spheres of the dumbbell, and the distance between them, characterised by the sphere separations $l$. Starting from a fluid, we have observed the formation of spherical and elongated micelles, vesicles and bilayers. In order to compare the outcome of the simulations 
    for our whole parameter space, we employ order parameters to identify and distinguish the clusters inside the simulation box. 
    
    To summarise, we have have investigated the effect of changing the sphere separation $l$ for three different size ratios $q=1.035$, $q= 1.25$ and $q= 0.95$, and for volume fractions ranging between $0$ and $0.25$.
    
    For size ratio $q=1.035$ we have observed the largest variety of structures. While micelles form for small sphere separations, and bilayers for high sphere separations, the formation of vesicles is limited to a very limited region of sphere separations. We speculate that this is due to a balance between interaction directionality and particle geometry and it would be interesting to study this further using free-energy calculations. 
    
    For size ratio $q=1.25$, upon increasing the sphere separation $l$, the structures change from spherical micelles to elongated micelles, and ultimately - at sufficiently high volume fractions - to bilayers. In this case, no vesicle formation was observed, suggesting that the system is very sensitive to small changes in particle geometry.
    
    Finally, for the last investigated case, where the size ratio $q=0.95$, we found bilayer formation on a wide range of sphere separations and volume fractions. We also found hollow structures such as vesicles and faceted polyhedra and occasionally droplet-like structures where the particles are clustered together with random particle orientations. Comparison with the state diagrams for $q=1.035$ and $q=1.25$ suggests that the size ratio is an important factor in stabilising the bilayers with respect to vesicles and micelles. 

In another paper on patchy Janus particles \cite{bib:cacciuto-janus} a transition is reported from liquid droplets to faceted polyhedra to bilayers, which is similar to what we observe for size ratio $q=0.95$ and small sphere separations, while Ref. \cite{bib:whitelam-peanuts} gives an estimate for the average cluster size consistent with ours, once we use similar geometric and interaction parameters.
    
This paper illustrates how a variety of different structures, some of them particularly relevant for applications \cite{bib:zheng-bilayer.photonics,bib:liu-amphiphile.nanostructures,bib:yang-wormlike.applications,bib:parak-colloidal.medicine} such as hollow vesicles and bilayers, can be formed starting from patchy dumbbells with attractive interactions. The work also shows that many of these structures can be well characterised using cluster order parameters. It also suggests that in experiments the structures might be very sensitive to variations in particle geometry, which is the case in for example, polydisperse systems. In particular, this information is useful when designing particles that can form hollow vesicles that can be useful in drug delivery or colloidal surfactants. A more detailed study focusing on this topic, ideally combined with experiments, might certainly be an interesting follow-up. Finally, in this paper we have focused on relatively low volume fractions where no crystallisation takes place. The formation of crystals, some of which might have interesting photonic properties, could also be topic of a follow-up study. 

\appendix
\section{}\label{apx:class.details}
\begin{figure*}[htb]
  \centering
  \includegraphics[scale=0.183]{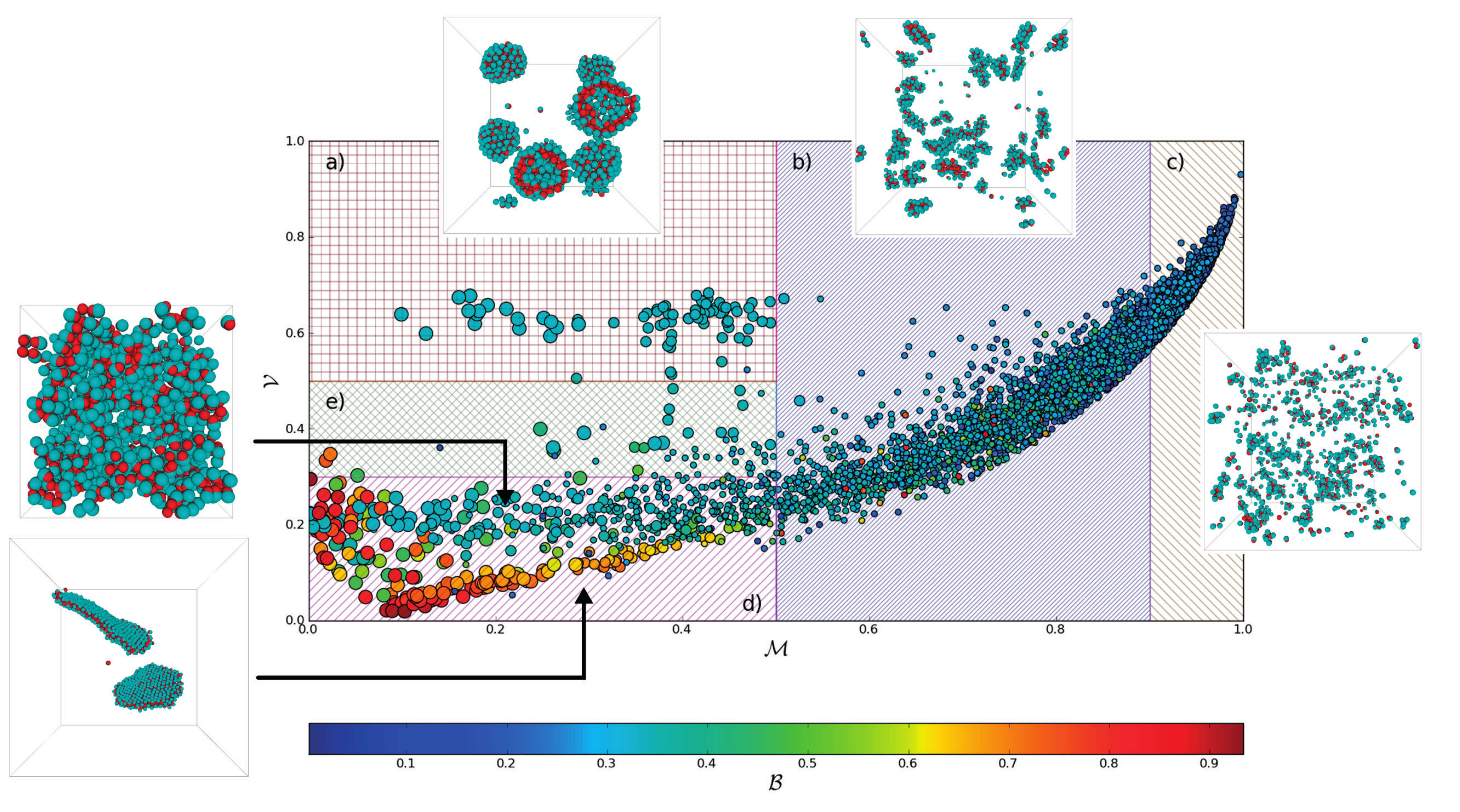}
  \caption{{\small Scatter plot of the values of the order parameters in the $(\mathcal{V},\mathcal{M})$ plane. Colour code and marker size stand for, respectively, the $\mathcal{B}$ parameter and the size of cluster (see  text). The following main structures can be identified: vesicles (region ``a''), elongated micelles (region ``b'' and ``d''), spherical micelles (region ``c''), and bilayers (region ``d''). Faceted polyhedra and liquid droplets fall both into region ``e'' of the diagram.}}
  \label{fig:mvnb-pr}
\end{figure*}
In order to identify and classify the self-assembled structures obtained for patchy dumbbells, we employ the order parameters $\mathcal{M},\mathcal{V}$ and $ \mathcal{B}$ as defined in Eq. \ref{eq:micelles}, \ref{eq:bilayers} and \ref{eq:vesicles}. In Fig. \ref{fig:mvnb-pr} we show the space of the order parameter values $(\mathcal{M},\mathcal{V},\mathcal{B})$ for clusters of patchy dumbbells for all the size ratios $q$, all the sphere separations $l$, and all the packing fractions $\phi$ that we considered in this study. Here, we consider all the results together for classification purposes. In the $x-y$ plane we have the $(\mathcal{V},\mathcal{M})$ parameters, while the $\mathcal{B}$ parameter is used to colour code the markers. Additionally, marker of three different sizes (small dots $\in [0,75]$, medium dots $\in [76,260]$, large dots $\in [261,1024]$) have been used to encode the cluster size information in the plot, in order to enhance its readability. We complement the plot in Fig. \ref{fig:mvnb-pr} with insets containing snapshots of the typical configurations in different regions of the diagram. We partition the diagram in different areas according to the values of the cluster order parameters, and we label them from ``a'' through ``e''. The three main regions in the diagram, form the vesicle (a), the micellar (b, c and partially d) and the bilayer regime (d). The micellar regime transforms continuously from elongated micelles at low $\mathcal{M}$ to more spherical micelles with increasing $\mathcal{M}$, and is identified by the main sequence in the diagram. Here, we have used a threshold value of $\mathcal{M}\geq 0.9$ to label aggregates as spherical micelles. The low-$\mathcal{V}$, low-$\mathcal{M}$ area (d) encloses large aggregates with high degree of orientational order ($\mathcal{B}$ parameter) -- the bilayer-like aggregates -- as well as smaller elongated micelles with low value of $\mathcal{B}$. Somewhat in between the vesicles and the bilayer regimes, liquid-like droplets as well as faceted polyhedra are observed (partition e). Note that, as Fig. \ref{fig:mvnb-pr} is a cumulative diagram, not all the structures are present in every size ratio considered, but rather each size ratio contributes differently to the areas in the diagram. While we could make as many diagrams as size ratios considered, we prefer to have a global and unique way of detecting different kind of aggregates.

    \begin{acknowledgments}
      \noindent
      This work is part of the research programme of the Foundation for Fundamental Research on Matter (FOM), which is part of the Netherlands Organisation for Scientific Research (NWO). T.V. and M.D. acknowledge financial support from a NWO-VICI grant. T.V. also acknowledges Marie Curie fellowship No. 623364 under the FP7-PEOPLE-2013-IEF program for financial support. G. Avvisati thanks the support from L. Taghizadeh, L. Avvisati, R. Perfetto, J. Elliott, B. Agarwal, A. Longobardi.
    \end{acknowledgments}

    \bibliography{references} %for local reference file
\end{document}